 \documentclass[final,3p]{elsarticle}

 \usepackage{graphics}
 \usepackage{graphicx}
 \usepackage{epsfig}

\usepackage{amssymb}
 \usepackage{amsthm}
\usepackage{amsmath}

\newlength{\dhatheight}

\makeatletter
\def\gsim{\compoundrel>\over\sim}

\def\compoundrel#1\over#2{\mathpalette\compoundreL{{#1}\over{#2}}}
\def\compoundreL#1#2{\compoundREL#1#2}
\def\compoundREL#1#2\over#3{\mathrel
         {\vcenter{\hbox{$\m@th\buildrel{#1#2}\over{#1#3}$}}}}
\makeatother

\journal{Astroparticle Physics}

\begin{document}

\begin{frontmatter}



 \title{A search for neutrino signal from dark matter annihilation in the center of the Milky Way with Baikal NT200}





\author[a]{A.D. Avrorin}
\author[a]{A.V. Avrorin}
\author[a]{V.M. Aynutdinov}
\author[g]{R. Bannasch}
\author[b]{I.A. Belolaptikov}
\author[c]{D.Yu. Bogorodsky}
\author[b]{V.B. Brudanin}
\author[c]{N.M. Budnev}
\author[a]{I.A. Danilchenko}
\author[a]{S.V. Demidov\corref{cor}}
\cortext[cor]{Corresponding authors}
\ead{demidov@ms2.inr.ac.ru}
\author[a]{G.V. Domogatsky}
\author[a]{A.A. Doroshenko}
\author[c]{A.N. Dyachok}
\author[a]{Zh.-A.M. Dzhilkibaev}
\author[e]{S.V. Fialkovsky}
\author[c]{A.R. Gafarov}
\author[a]{O.N. Gaponenko}
\author[a]{K.V. Golubkov}
\author[c]{T.I. Gress}
\author[b]{Z. Honz}
\author[g]{K.G. Kebkal}
\author[g]{O.G. Kebkal}
\author[b]{K.V. Konischev}
\author[c]{A.V. Korobchenko}
\author[a]{A.P. Koshechkin}
\author[a]{F.K. Koshel}
\author[d]{A.V. Kozhin}
\author[e]{V.F. Kulepov}
\author[a]{D.A. Kuleshov}
\author[a]{V.I. Ljashuk}
\author[e]{M.B. Milenin}
\author[c]{R.A. Mirgazov}
\author[d]{E.R. Osipova}
\author[a]{A.I. Panfilov}
\author[c]{L.V. Pan'kov}
\author[b]{E.N. Pliskovsky}
\author[f]{M.I. Rozanov}
\author[c]{E.V. Rjabov}
\author[b]{B.A. Shaybonov}
\author[a]{A.A. Sheifler}
\author[a]{M.D. Shelepov}
\author[d]{A.V. Skurihin}
\author[b]{A.A. Smagina}
\author[a]{O.V. Suvorova\corref{cor}}
\ead{suvorova@cpc.inr.ac.ru}
\author[c]{V.A. Tabolenko}
\author[c]{B.A. Tarashansky}
\author[g]{S.A. Yakovlev}
\author[c]{A.V. Zagorodnikov}
\author[a]{V.A. Zhukov}
\author[c]{V.L. Zurbanov}

\address[a]{Institute for Nuclear Research, 60th October Anniversary pr. 7A, Moscow 117312, Russia}
\address[b]{Joint Institute for Nuclear Research, Dubna 141980, Russia}
\address[c]{ Irkutsk State University, Irkutsk 664003, Russia}
\address[d]{ Skobeltsyn Institute of Nuclear Physics  MSU, Moscow 119991, Russia}
\address[e]{ Nizhni Novgorod State Technical University, Nizhni Novgorod 603950, Russia}
\address[f]{ St. Petersburg State Marine University, St. Petersburg
  190008, Russia} 
\address[g]{ EvoLogics GmbH, Berlin, Germany}

\begin{abstract}
We reanalyze the dataset collected during the years 1998--2003 by the
deep underwater neutrino telescope NT200 in the lake Baikal with the
low energy threshold (10 GeV) in searches for neutrino signal from
dark matter annihilations near the center of the Milky Way. Two
different approaches are used in the present analysis: counting events
in the cones around the direction towards the Galactic Center and the
maximum likelihood method. We assume that the dark matter particles
annihilate dominantly over one of the annihilation channels
$b\bar{b}$, $W^+W^-$, $\tau^+\tau^-$, $\mu^+\mu^-$ or
$\nu\bar{\nu}$. No significant excess of events towards the Galactic
Center over expected neutrino background of atmospheric origin is
found and we derive 90\% CL upper limits on the annihilation cross
section of dark matter.
\end{abstract}


\end{frontmatter}

\section{Introduction}
Today all cosmological and astrophysical observations are successfully
explained within a paradigm of the standard cosmological model
($\Lambda$CDM) stating that the most of the energy density of
the Universe is stored in the dark energy or cosmological constant
($\Lambda$, about 68\%) and non-baryonic cold dark matter (CDM, about
27\%). Unambiguous presence of the latter component is confirmed 
by measurements of galaxy rotation curves~\cite{Persic:1995ru}, weak
gravitational lensing of distant objects like galaxy
clusters~\cite{Abel520}, measurements of properties of cosmic
microwave background~\cite{WMAP2003,WMAP2012,Planck2013}, analysis of
structure formation~\cite{Primack:1997av} and
nucleosynthesis~\cite{Jedamzik:2009uy} (see also
Ref.~\cite{Bertone:2004pz} for a review).

One of the most favorable ideas for explaining dark matter (DM)
phenomena is Weakly Interacting Massive Particles or
WIMPs~\cite{Steigman:1984ac}. In this scenario predicted by several
classes of models of new physics~\cite{Jungman:1995df,Hooper:2007qk},
these particles are supposed to be in thermal equilibrium in the early
Universe and can annihilate into Standard Model particles. But as the
Universe was expanding and cooling down, the annihilation processes
ceased out and number density of the dark matter particles became
frozen out at some value which is determined by their
annihilation cross section. Thus, at least in the WIMP scenario
one can expect some signal from the dark matter annihilations towards 
  the directions of local overpopulation of these particles. The
Galactic Center (GC) of the Milky Way is one such direction. 

Two types of messengers from DM annihilation signal in the Galactic
Center, i.e. gamma rays and neutrinos, are expected to be
detected by the telescopes. They both originate in the same
energy ranges (GeV--TeV for WIMPs) in decays of particles produced in
kinematically allowed dark matter annihilation channels. Several
analyses of diffuse gamma-rays from the FERMI-LAT dataset (pass 7)
point out on an evidence for central and spatially extended excess
toward the Galactic Center (GCE).
Large astrophysical uncertainties to the background of diffuse
  gamma-radiation in the Galaxy generate ambiguous interpretations of
  the GCE. Among the scenarios explaining the GCE there have been
  discussed possibilities of unresolved conventional astrophysical
  gamma-ray sources, e.g. millisecond pulsars~\cite{Abazajian:2014fta}
  as well as different extensions of the Standard Model with dark matter 
  particles annihilating in the MW halo~\cite{Hooper:2014} (for
  a review see e.g. Ref.~\cite{Conrad:2015bsa}).
The analysis performed in Ref.~\cite{Hooper:2014} which uses
  filtered emissions from individual point sources like globular
  clusters and millisecond pulsars detected by the
  FERMI-LAT,
supports a DM signal from annihilations either into $b\bar{b}$ channel 
with dark matter particle of the mass about 30--60 GeV or into
$\tau^{+}\tau^{-}$ channel for the masses in 5--15 GeV range. The
annihilation cross section is found 
to be about $10^{-26} {\rm cm}^3/{\rm s}$ which is close to the
value predicted in the scenario of WIMPs annihilation in the early
Universe. The latest analysis of the FERMI-LAT dataset (pass
8)~\cite{FERMI:2015p8} 
shows some tension in consistency with the dark matter
interpretations  of the GCE, related to the dark matter density
profile and the value of local dark matter density. 

The operating neutrino telescopes have not yet observed a signal
from the dark matter annihilations in the GC over
expected atmospheric neutrino background, see recent analyses from
the ANTARES~\cite{Adrian-Martinez:2015wey}, IceCube~\cite{Aartsen:2015xej} 
and Super-Kamiokande~\cite{Frankiewicz:2015zma} collaborations. In this paper we
analyze the dataset of upward going 
muons produced by neutrinos in the lake Baikal and measured with 
  the NT200 
telescope for a period between April of 1998 to February of 2003
in search for an excess in the directions near the GC. We
focus on the dark matter particles with masses from 30~GeV to 10~TeV
annihilating 
dominantly into one of the following benchmark annihilation
channels: $b\bar{b}$, $W^+W^-$, $\tau^+\tau^-$, $\mu^+\mu^-$ and also
flavor-symmetric neutrino channel $\nu\bar{\nu}$. Two independent
analyses of the 
data are performed and consistent results are found. Finally, we
obtain upper limits on the dark matter annihilation cross section for
these annihilation channels. Also we discuss influence of systematic
uncertainties on the obtained results.

\section{Experiment and data sample}
\label{sec:experiment}
\begin{figure}
\begin{center}
\includegraphics[width=0.5\textwidth]{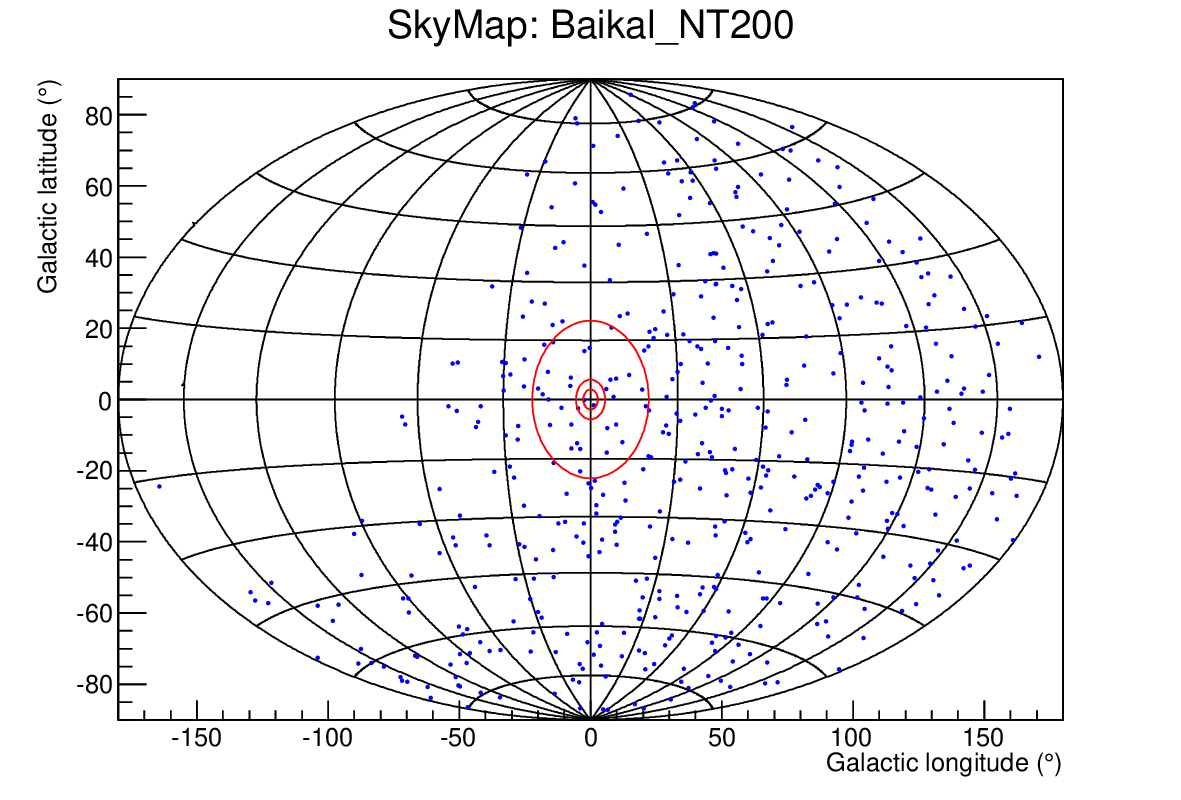}
\end{center}
\caption{\label{fig:GlxSky} 
The NT200 sample of neutrino events in galactic coordinates}  
\end{figure}

The NT200 is a deep underwater neutrino telescope in lake Baikal 
which began data taking in 1993. This detector was 
the first that proved the method of study of high energy muons,
  which come
from top or bottom hemisphere across a large volume of natural water, 
by recording their Cherenkov radiation~\cite{NT200-1997}. 
Lake Baikal deep water is characterized by the absorption length of 
$L_{abs}(480~{\rm nm})=20-24$~m, scattering length of $L_{sc}=$30--70 m
and strongly anisotropic scattering function with the mean
cosine of the 
scattering angle of 0.85--0.9. The Cherenkov light of relativistic particles
is recorded at appropriate wavelengths by an array of optical modules 
(OMs) which are time-synchronized and energy-calibrated by artificial
light pulses. At 1 km water depth, the muon flux from cosmic ray
interactions in the upper atmosphere is about one million times higher
than the flux of upward going muons initiated by neutrino interactions
in the water and rock below the array. Selection of clean
neutrino event 
sample of true upward  going muons is a major challenge which requires
highly  efficient rejection of misreconstructed downward moving muons.  
\begin{figure}[!htb]
\begin{center}
\begin{tabular}{cc}
\includegraphics[width=0.45\textwidth,angle=0]{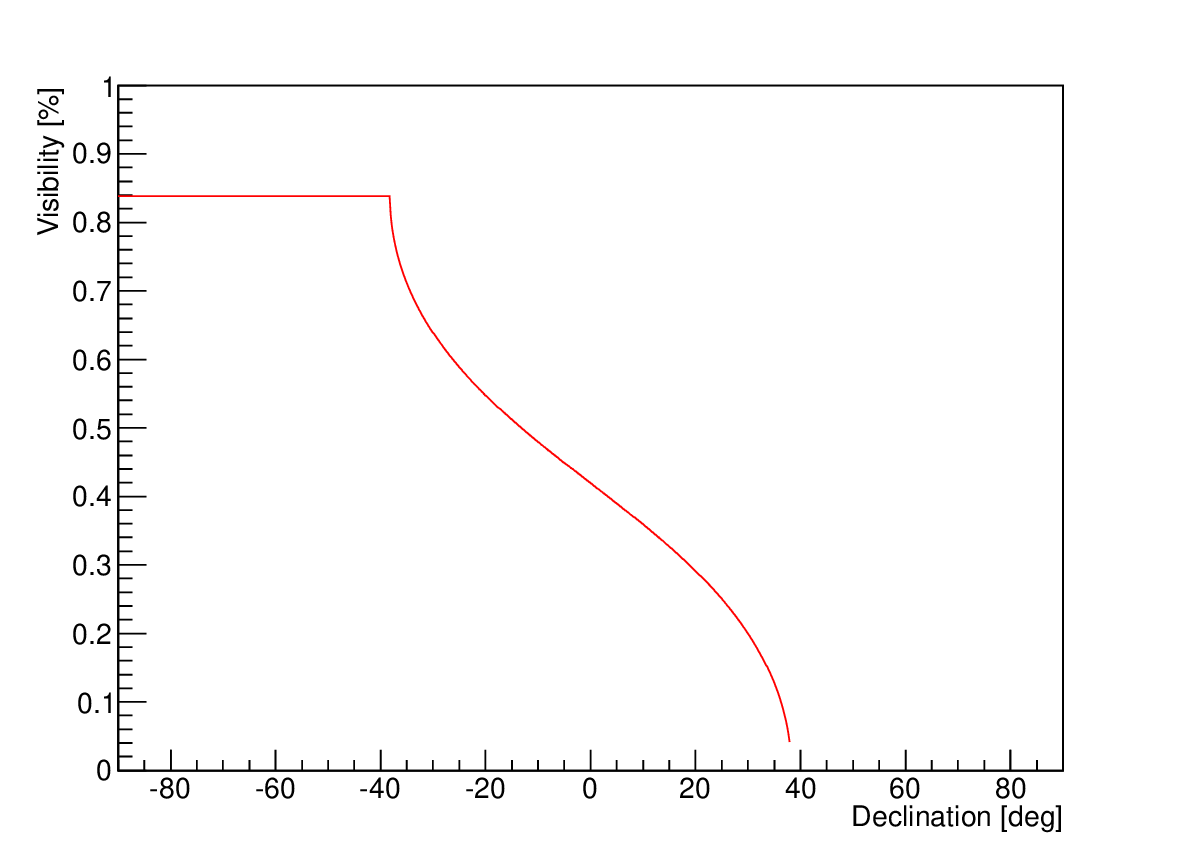}
\includegraphics[width=0.45\textwidth,angle=0]{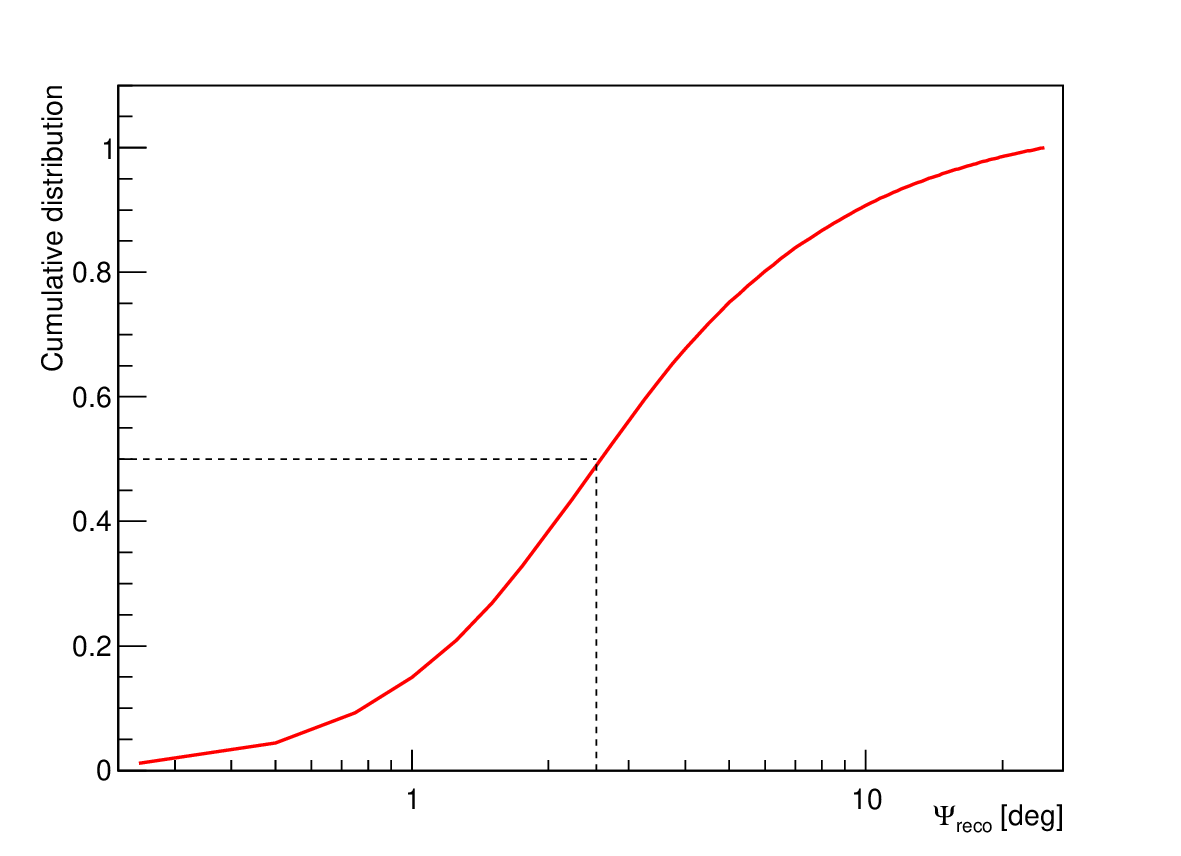}
\end{tabular}
\end{center}
\caption{\label{fig:Vis}  
Left: The NT200 visibility as a function of the declination.  
Right: The cumulative distribution of the reconstruction angle $\Psi_{reco}$
for the NT200 data analysis. The black-dotted line indicates the median
value. }
\end{figure}
The NT200 instrumentation volume encloses 100 Ktons placed
in the southern basin of the lake Baikal, at the distance of 3.5
km off the shore and at the depth of \mbox{1.1 km}. In
Fig.~\ref{fig:Vis} (on left), the 
visibility over declinations for the NT200 site located at
  51.83$^\circ$  of Northern latitude is shown. Here we account 
  for its dead time when the detector had to be upgraded in the
winter expeditions~\cite{NT200-2006}. Note, that in April 2015 
there has been deployed a new larger detector~\cite{GVD2015a}
operating now at this place as a demonstration cluster of
about 2 Mton size for a future Gigaton volume
detector~\cite{GVD2015b}. Sensitivity of the future experiment to 
  the DM annihilation signal from the GC has been discussed in
Ref.~\cite{GC-GVD2015}. 
The NT200 configurations, functional systems, calibration
methods and software for  muon track reconstruction have been
described elsewhere ~\cite{NT200-2006, Baikal2007, NT200Mono-2008,
  Baikal2009, NT200Gamma-2011, NT200Astro-2011}. The detector consists 
of 192 optical modules arranged pairwise on 8 strings of 68.75\,m
length: seven peripheral strings and a central one. The distance
between the nearest strings is 21.5\,m. Each OM contains hybrid
photodetector QUASAR-370, a photo multiplier tube (PMT) with 37-cm
diameter. To suppress background from dark noise, two PMTs of a
pair are switched in coincidence within the time window of
15\,ns.   
Present analysis is based on the data collected between April of
1998 and 
February of 2003, with in total 2.76 live years, and taken with the
muon trigger. The trigger requires \mbox{N$_{hit} \geq n$}
within \mbox{500\,ns}, where hit refers to a pair of fired
OMs coupled in a {\it channel}. Typically the value of \mbox{n}
is set to \mbox{3 or 4}. We use the same dataset and Monte Carlo
  (MC) sample as in Ref.~\cite{NT200Sun:2014swy}. The detector
response to the atmospheric muons and neutrinos has been
obtained with 
MC simulations based on standard codes
{\texttt{CORSIKA}}~\cite{CORSIKA} and {\texttt{MUM }}\cite{MUM} 
  using the Bartol atmospheric $\nu$ flux~\cite{Bartol}. To
distinguish upward and downward going muons on 
one-per-million mis-assignment level, a filter with several levels of
quality cuts was developed for the atmospheric neutrino ($\nu_{atm}$)
analysis \cite{Belolap07}. The atmospheric muons which have been
mis-reconstructed as upward-going particles are the main source 
of the background in the search for neutrino induced upward-going
muons. The offline filter which requires at least 6 hits on at least 3
strings ("6/3") selects about $40\%$ of all triggered events. At this
level the r.m.s. mismatch angle $\psi_{reco}$ between the direction of
 incoming muon and its reconstructed value is about $14.1^\circ$
for the $\nu_{atm}$-sample. To get the best possible estimator for the 
direction, we use multiple start guesses for the $\chi^{2}$
minimization~\cite{NT200Astro-2011}. For the final choice
of the local minimum of $\chi^{2}$, we use  quality parameters which
are not related to the time information. The quality cuts are applied
to variables like the number of hit channels, $\chi^2/d.o.f.$, the
probability of fired channels to have been hit or not and the actual
position of the track with respect to the detector center. To improve
the signal-to-background ratio we use only events with the
reconstructed zenith angle $\Theta > 100^\circ$. All the cuts provide
rejection factor for the atmospheric muons of about $10^{-7}$,
resulting in the neutrino energy threshold of about 10~GeV, dispersion
of $4.5^\circ$ for the distribution of mismatch angles and the
median value $2.5^\circ$ as it is seen in Fig.~\ref{fig:Vis}
  (right) (for more details see
  Ref.~\cite{NT200Astro-2011,Belolap07}).  
In Fig.~\ref{fig:GlxSky} we show the arrival directions of
reconstructed muons in galactic coordinates and cones around the GC
with opening angles 20$^\circ$, 5$^\circ$ and 2.5$^\circ$ 
  containing 31, 2 and 2 observed events, respectively.
In the next Section, we discuss expected signal from the dark matter
  annihilations in our Galaxy and background from
  the atmospheric neutrinos. 

\section{Signal and background}
\begin{figure}
\begin{center}
\includegraphics[width=0.35\textwidth,angle=-90]{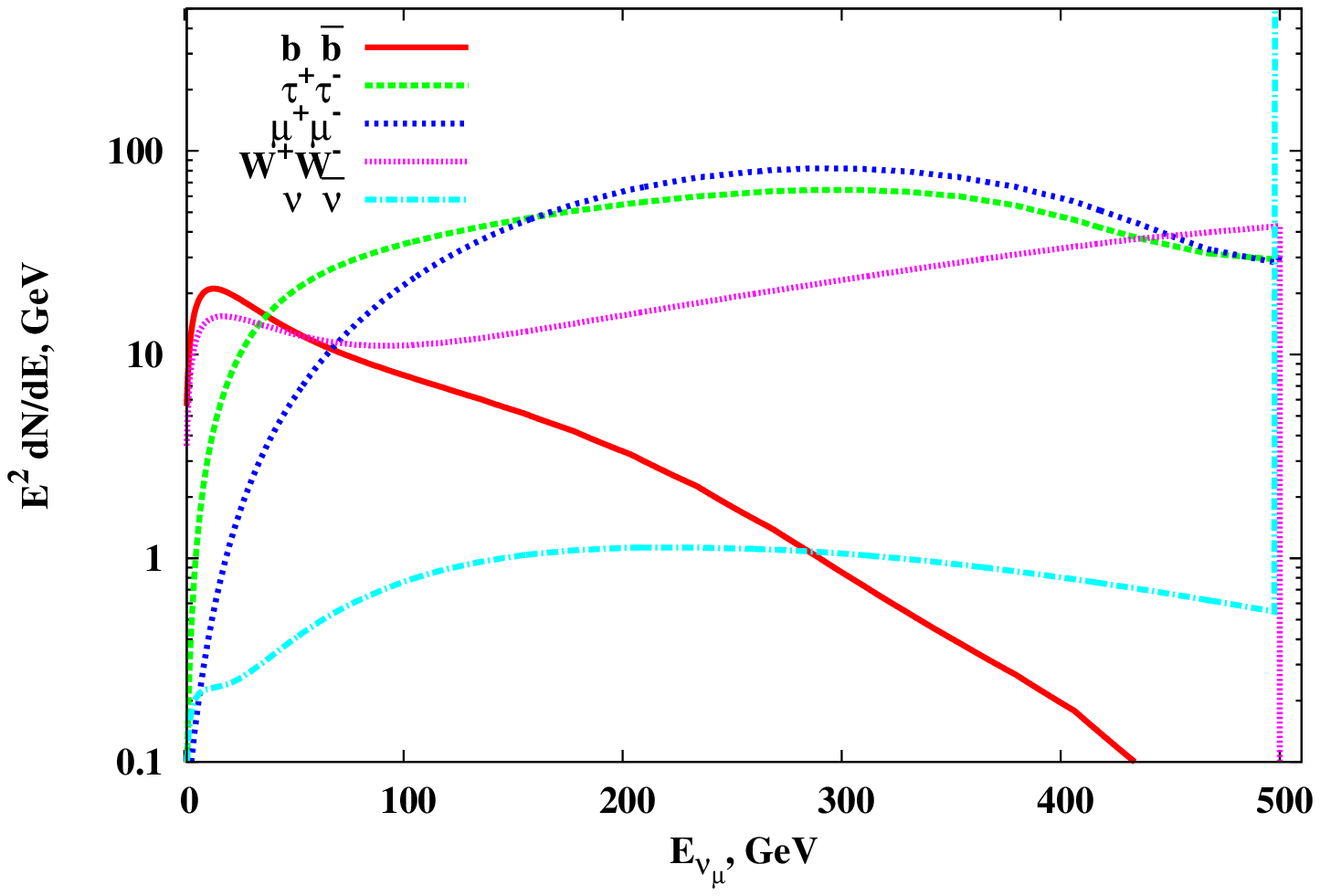}
\end{center}
\caption{\label{fig:2} 
Neutrino $\nu_{\mu}$ energy spectra at the Earth for
$m_{DM}=500$~GeV.} 
\end{figure}

Neutrino flux at the Earth from the dark matter annihilations in the
Galactic Center from a particular direction has the following
form 
\begin{equation}
\label{eq:1}
\frac{d\phi_{\nu}}{dE_{\nu}d\Omega} = \frac{\langle \sigma_a v\rangle}{2}\;
J_{a}(\psi)\;\frac{R_0\rho_{0}^{2}}{4\pi m_{DM}^2}\;\frac{dN_{\nu}}{dE_{\nu}},
\end{equation}
where $\langle \sigma_a v\rangle$ is the annihilation cross
section averaged over velocity distribution of the dark matter particles
at present epoch and $dN_{\nu}/dE_{\nu}$ is the (anti)neutrino
energy spectrum. The astrophysical factor $J_a(\psi)$ can be
parametrized as 
a function of angular distance $\psi$ from the GC to the chosen
direction 
and has the form 
\begin{equation}
\label{eq:2}
J_a(\psi) = \int_{0}^{l_{max}}\frac{dl}{R_0}\;
\frac{\rho^2\left(\sqrt{R_0^2-2rR_0\cos{\psi}+r^2}\right)}{\rho_{0}^2}.
\end{equation}
Here $R_0$ is the distance from the GC to the Solar System, $\rho_0$ is
the local dark matter density and integration goes along line-of-sight
till $l_{max}$ which is much larger than the size of the Milky
Way. Several different models are used to describe the dark matter
density distribution of our Galaxy, see
e.g. Refs.~\cite{Navarro:1995iw,Navarro:1996gj,Kravtsov:1997dp,Moore:1999nt,Burkert:1995yz}. 
Numerical N-body simulations~\cite{Kuhlen:2012ft} show that dark
matter form an almost spherical halo. Simulations without baryons
predict cuspy profiles. However, inclusion of the baryons can 
change these 
conclusions and N-body simulations being limited in number of
particles can not resolve small area around the center of galaxy, so
the main uncertainty in the profile comes from this region. At the
same time the signal from annihilations is proportional to the
square of the dark matter density, see Eq.~(\ref{eq:1}). So,
this part of astrophysical input is the main theoretical uncertainty
for the signal from DM
annihilations in the Galaxy. Direct observational data of the Milky Way
can not resolve this uncertainty, because the part of our Galaxy
within the Solar System circle is dominated by baryons, so the influence 
of the dark matter within this circle on motion of astrophysical
objects is small. Even the local dark matter density is known with a
large uncertainty 0.2--0.6~GeV$\cdot$cm$^{-3}$. Moreover, one can not exclude a
possibility that dark matter can form clumps in our Galaxy and that
the signal from a particular directions can be increased by the presence
of a clump along line-of-sight~\cite{Berezinsky:2014wya}. We will
present the final results for
Navarro-Frenk-White~(NFW)~\cite{Navarro:1995iw,Navarro:1996gj} model of the DM density profile
and compare them with more cuspy profile in the Moore~\cite{Moore:1999nt} model and
cored profile of the Burkert~\cite{Burkert:1995yz} model. The latter one is a
currently favored by the observational data~\cite{Nesti:2013uwa}.
All these profiles can be described as follows  

\begin{equation}
\label{eq:3}
\rho(r) = \frac{\rho_*}{\left(\delta + \frac{r}{r_*}\right)^{\gamma}\left[1 +
    \left(\frac{r}{r_*}\right)^{\alpha}\right]^{(\beta-\gamma)/\alpha}}
\end{equation}
with parameters presented in Table~\ref{tab:1}.
\begin{table}
\begin{center}
\begin{tabular}{|c|c|c|c|c|c|c|}
\hline
Model & $\alpha$ & $\beta$ & $\gamma$ & $\delta$ & $r_*$, kpc &
$\rho_*$, GeV/cm$^3$ \\
\hline
NFW  & 1 & 3 & 1 & 0 & 20 & 0.3 \\
\hline
Burkert & 2 & 3 & 1 & 1 & 9.26 & 1.88\\
\hline
Moore & 1.5 & 3 & 1.5 & 0 & 28 & 0.27 \\
\hline
\end{tabular}
\caption{\label{tab:1} Parameters of DM density
    profiles.}  
\end{center}
\end{table}
The neutrino energy spectra entering Eq.~(\ref{eq:1}) are
determined by 
a particular theoretical model. In a model independent approach followed
in the present study, it is assumed that the dark matter
particles annihilate over particular annihilation channel with 100\%
branching ratio. We use $b\bar{b}$, $\tau^{+}\tau^{-}$,
$\mu^{+}\mu^{-}$, $W^{+}W^{-}$ channels as well as $\nu\bar{\nu}\equiv 
\frac{1}{3}(\nu_e\bar{\nu}_e + \nu_\mu\bar{\nu}_\mu +
\nu_\tau\bar{\nu_\tau})$ which is flavor symmetric combination of 
(almost) monochromatic neutrino. We consider the masses of the dark matter
particles from 30~GeV to 10~TeV. For the energy spectra of these 
annihilation channels we use the results of
Ref.~\cite{Baratella:2013fya} which include electroweak corrections 
important for large masses of the particles. After propagation
over astrophysically large distances, neutrinos from the GC arrive at
the Earth as mass states and we use the following
set~\cite{Forero:2014bxa} of the oscillation parameters: $\Delta
m_{21}^2 
= 7.6\cdot 10^{-5}~{\rm eV}^2$, $\Delta m_{31}^{2} = 2.48\cdot
10^{-3}~{\rm eV}^2$, $\delta_{CP} = 0$, $\sin^2{\theta_{12}} = 0.323$,
$\sin^2{\theta_{23}} = 0.567$, $\sin^2{\theta_{13}} = 0.0234$ to
calculate the muon (anti)neutrino energy spectrum. We simulate
neutrino 
propagation through the Earth to the detector level as described
in~\cite{Boliev:2013ai}. Final neutrino energy spectra are presented
in Fig.~\ref{fig:2} as an example for $m_{DM}=500$~GeV\footnote{
For the case of $\nu\bar{\nu}$ channel, an unphysical width was
introduced in Ref.~\cite{Baratella:2013fya}. We changed these spectra
back to their physical width conserving their normalization.
}. 

\begin{figure}[!htb]
\begin{center}
\includegraphics[width=0.35\textwidth,angle=-90]{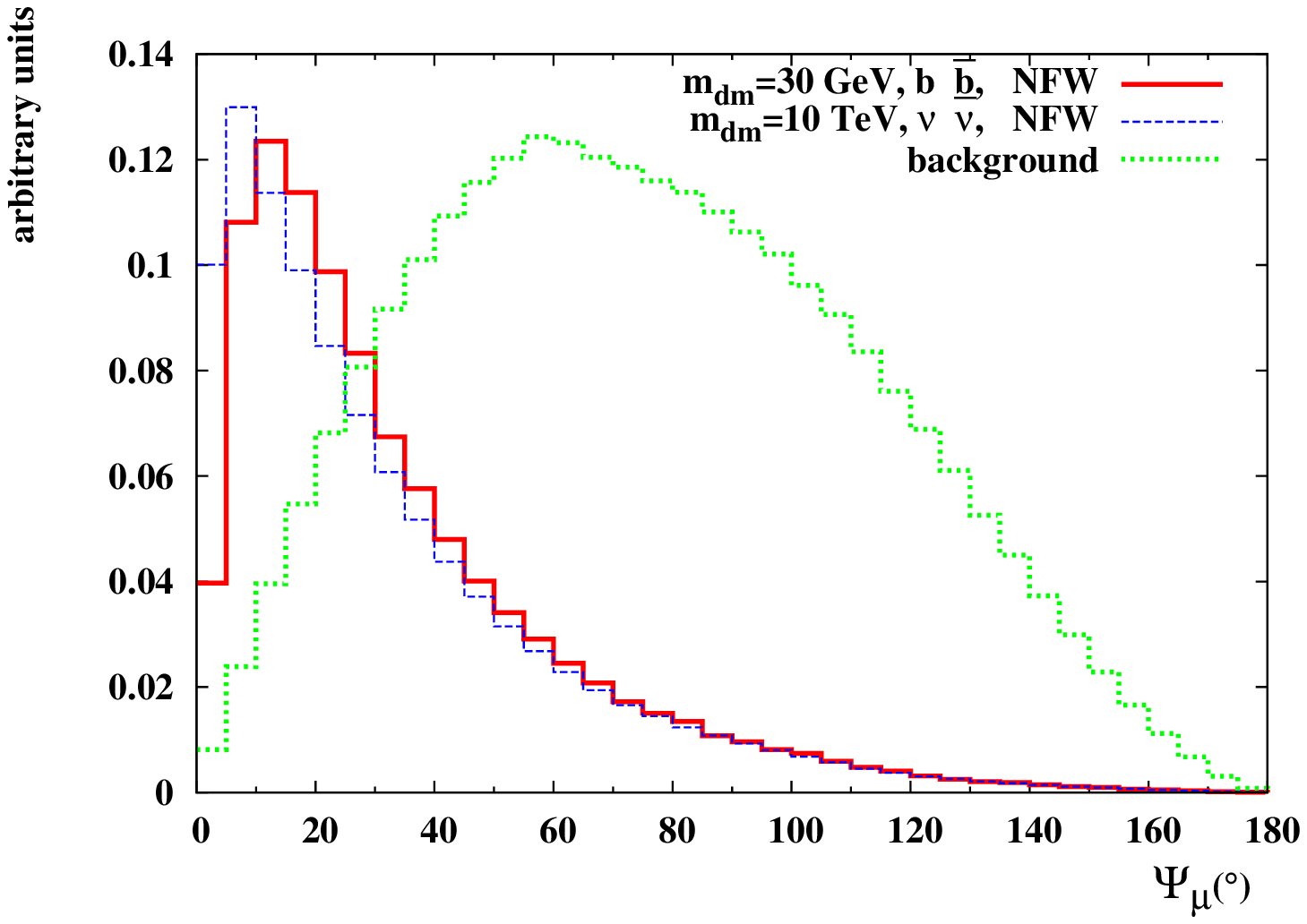}
\end{center}
\caption{\label{fig:4} 
Expected reconstructed signal and background angular distributions.}  
\end{figure}

Expected angular distributions of reconstructed signal events (muons), 
was obtained by MC simulations. The angular spread of the
  signal is determined by the behaviour 
  of the astrophysical factor $J_a(\psi)$ for a DM density profile,
  angular distribution of 
  muons from CC interactions of neutrinos for a particular
  annihilation channel as well as angular   resolution of the
  telescope. The latter has been taken into account 
  by additionally smearing the signal according to the
  distribution of mismatch angles obtained with MC and whose
    cumulative distribution is shown in Fig.~\ref{fig:Vis} (on right)
    (instead of 
  Gaussian distribution with dispersion equal to the angular
  resolution). 
Expected reconstructed angular distributions of the signal are
presented in Fig.~\ref{fig:4}   
for two opposite cases, the softest ($b\bar{b}$, $m_{DM}=30$~GeV) and
the hardest ($\nu\bar{\nu}$, $m_{DM}=10$~TeV) neutrino energy spectra
for NFW density profile.

\begin{figure}[!htb]
\begin{center}
\includegraphics[width=0.32\textwidth,angle=-90]{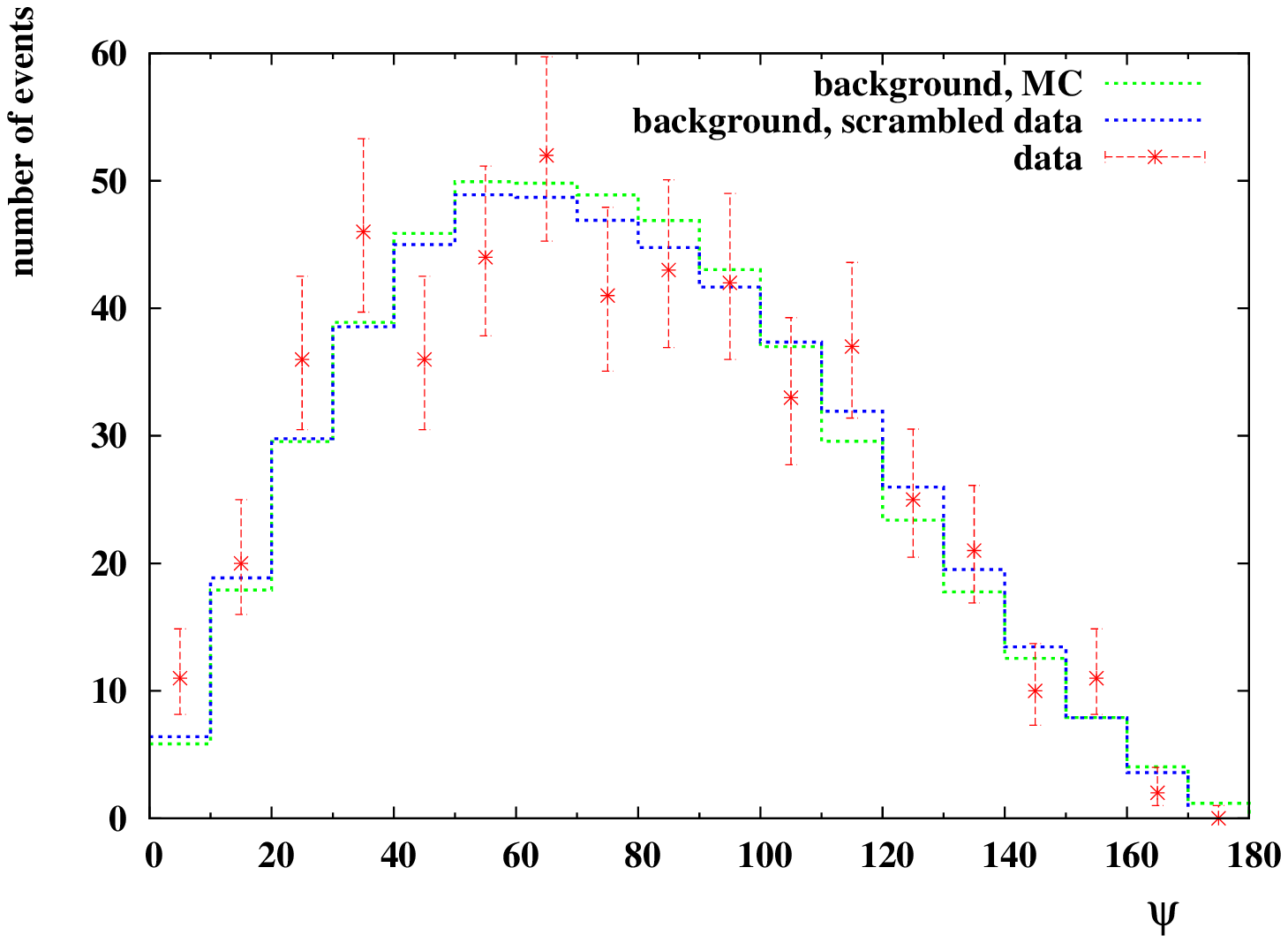}
\end{center}
\caption{\label{fig:3} 
The data with statistical errors (red) in comparison with 
  the background obtained using scrambling of the data (blue)
and from Monte-Carlo (green).}   
\end{figure}

The background for the process in question is dominated by
upgoing atmospheric neutrinos. To avoid large systematic errors,
typically resulting from MC simulations, in the following
analysis we use the expected background which is estimated from the
data using their scrambling by randomization of right ascension of the
events. For scrambling, we use the same full data sample of
  reconstructed neutrino events which has been selected with the cuts
  discussed in Sec.2. 
The form of the background obtained in this way is shown in
Fig.~\ref{fig:3} by blue 
histogram. In addition to the scrambling, we introduce a correction to 
overall normalization of this background distribution by fitting it
with the real data outside the region of expected signal
contamination: we do it for $\psi > 60^{\circ}$. 
It is clear that using the full set of the data in this procedure
  we can not exclude a possible small signal contamination in our
  determination of the background.
For cross checking,
we use another procedure to obtain the shape of the
background angular distribution. Namely, we simulate the reconstructed
atmospheric neutrino angular distribution taking into account the
visibility which is the part of observation time of a point
  on the sky depending on its declination. By using the visibility we
  simulate angular distribution of the background and impose the same 
  cut on zenith angle to be less than 100 degrees as for real data. 
Again, we fit
obtained shape with the data in the region $\psi >
60^{\circ}$. The comparison between two different background 
  models is presented in Fig.~\ref{fig:3} along with the data angular
distribution. 

By comparison of obtained angular distributions of signal and
background with the data presented in Figs.~\ref{fig:3}
and~\ref{fig:4}, we see that there is a small excess in number of
observed events toward the GC. 
Below we estimate statistical significance of the excess and obtain
upper limits on number of signal events for each particular DM mass
and annihilation channel. Concerning the cones shown in
Fig.~\ref{fig:GlxSky}, the numbers of the expected
background events (observed events) inside them are 
  25.1 (31), 1.63 (2) and 0.42 (2).  

\section{Data analyses}
In this section we describe two different methods to analyze the data
and to look for neutrino signal from the dark matter annihilations
in the GC. Expected signal and background have
different energy and angular distribution. Here we can use only
angular information of the reconstructed events.  

\subsection{Method A: Optimization of the cone size}

\begin{figure}[!htb]
\begin{center}
\includegraphics[width=0.35\textwidth,angle=-90]{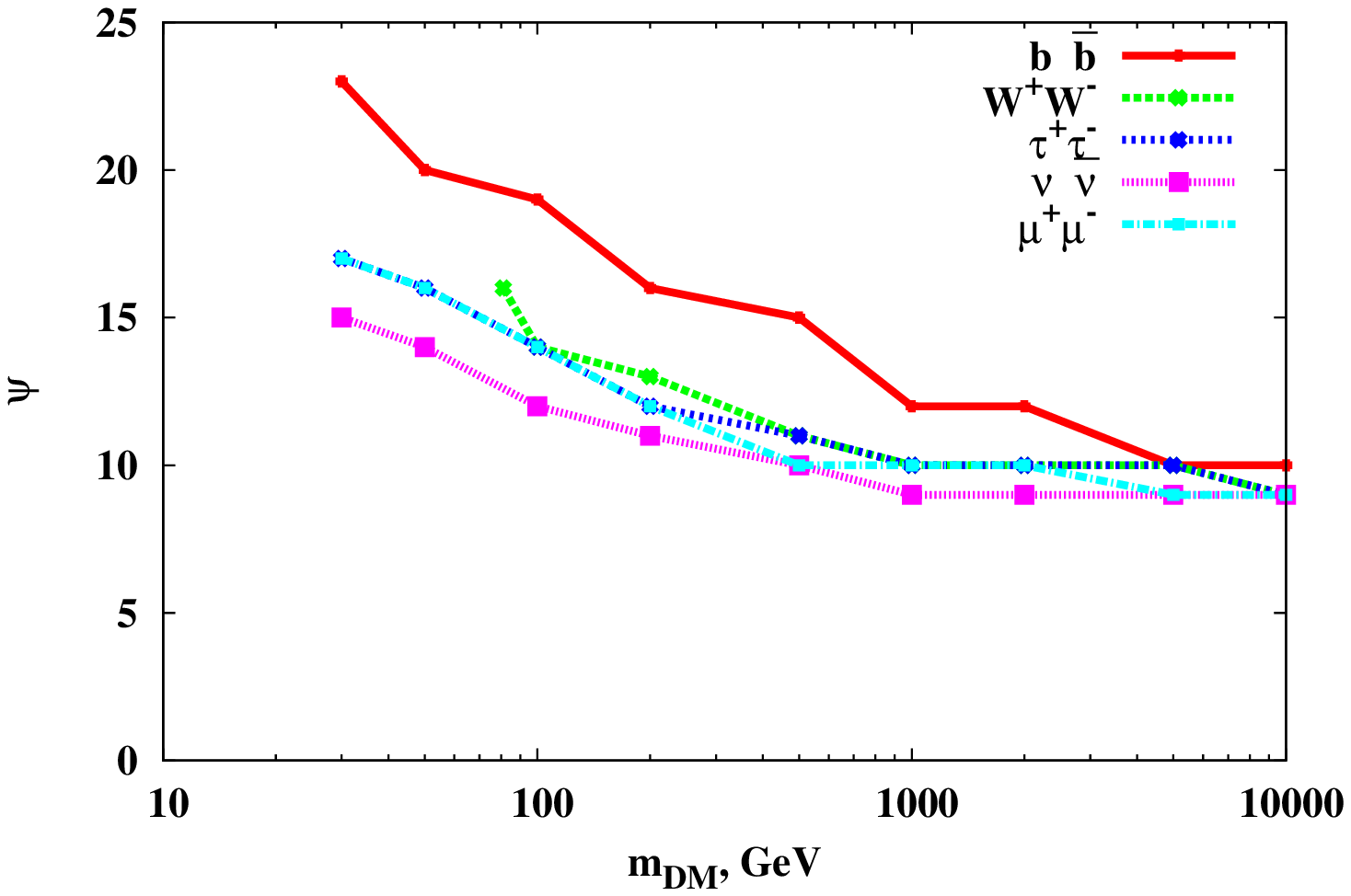}
\end{center}
\caption{\label{fig:5} 
The optimal values of cone half-angles for different dark matter
masses and annihilation channels.} 
\end{figure}

For analysis {\it A}, we choose a cone around the direction towards
the GC with half-open angle $\Psi$ and thus obtain a counting
experiment with expected number of signal $N_S(\Psi)$ and background
$N_B(\Psi)$ as well as observed $N_{obs}(\Psi)$ number of events. The
size of the cone is optimized by choosing maximal value for
signal-to-noise ratio. For blindness of the analysis we do it
  without use of the data in the search region. Namely,
  following the MRF 
approach~\cite{Hill:2002nv} we construct the quantity 
\begin{equation}
\frac{S}{N} \equiv \frac{\bar{N}_{S}^{90}(\Psi)}{\sqrt{N_B(\Psi)}},
\end{equation}
where $\bar{N}_{S}^{90}(\Psi)$ is 90\% CL upper limit on the number of
signal neutrino events inside given cone of the size $\Psi$ averaged
over number of observed events with Poisson distribution under 
background only hypothesis. The optimal
values of $\Psi$ for different dark matter masses and annihilation
channels are presented in Fig.~\ref{fig:5}.
They vary from 9$^{\circ}$ for channels with hard $\nu$
spectra to about 23$^{\circ}$ for annihilations with soft
neutrinos. Obtained large values are due to the considerable
angular spread of neutrino signal (for the NFW profile) as well due to
the broad distribution of the reconstructed mismatch angles.   

\begin{figure}[!htb]
\begin{center}
\includegraphics[width=0.35\textwidth,angle=-90]{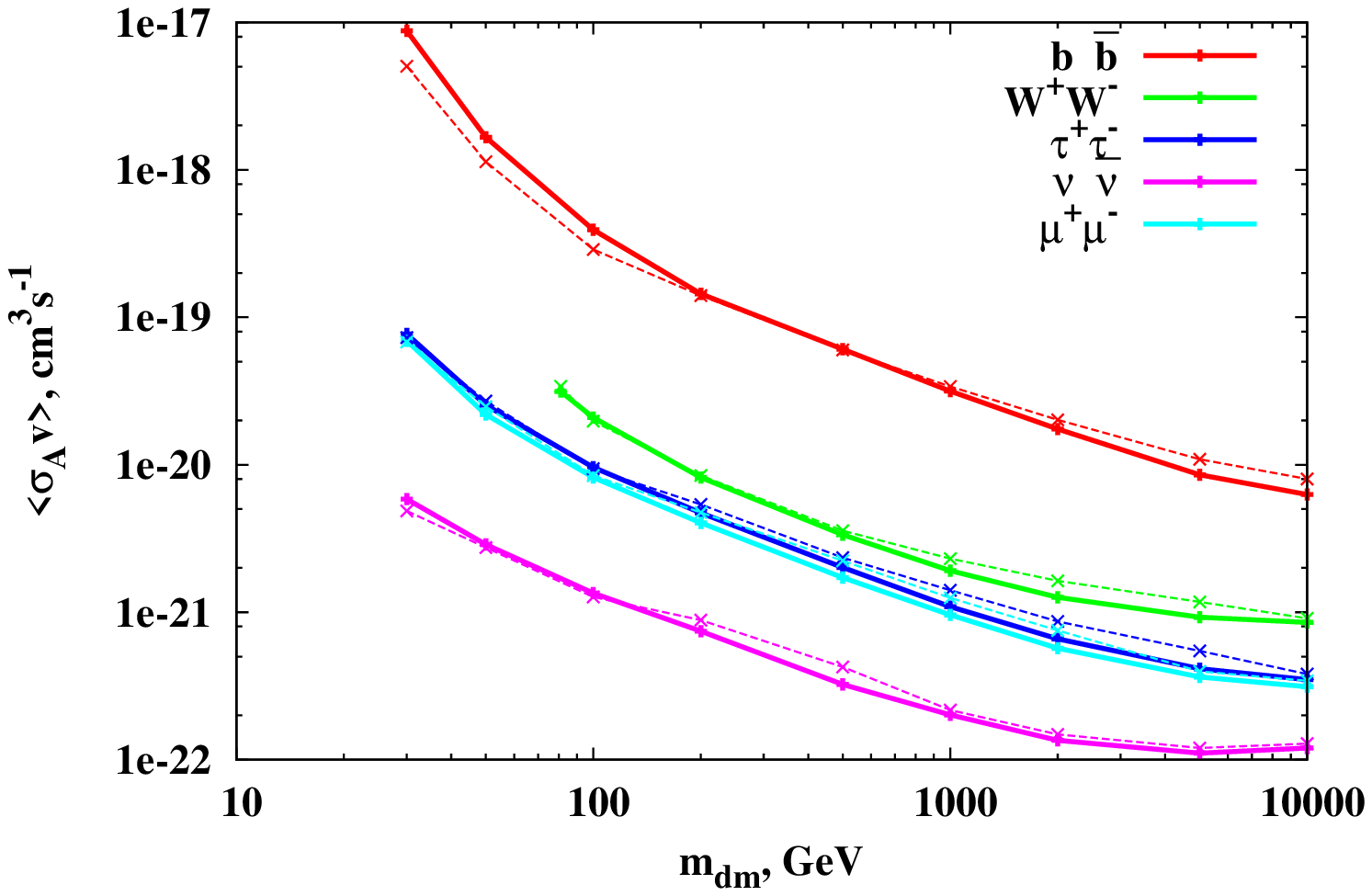}
\end{center}
\caption{\label{fig:11} 
90\% CL upper limits on the annihilation cross section for
methods {\it A} (dashed) and {\it B} (solid) assuming NFW density
profile.}   
\end{figure}

Inside the cones of optimal size, we set upper limits on the
number of signal events $N_S^{90}$ for a given DM mass and
annihilation channel by using expected background distribution toward
the GC. We apply TRolke class~\cite{trolke} in ROOT~\cite{ROOT} to get
the numbers taking into account systematic uncertainties to be
discussed in the next Section. These upper limits are transformed into
the upper limits on the dark 
matter annihilation cross section by using the expression for expected  
number of signal events from the GC direction for a live time $T$
of observation inside  a given cone 
\begin{equation}
\label{eq:6}
N(\Psi) = T\frac{\langle\sigma_a v\rangle R_0\rho_0^{2}}{8\pi
  m_{DM}^2}J_{a, \Delta\Omega} S^{eff} \int_{E_{th}}^{m_{DM}} dE \frac{dN_{\nu}}{dE_{\nu}}.
\end{equation}
Here $S^{eff}$ is the effective area of the telescope averaged
over neutrino energy spectrum for a particular annihilation channel 
\begin{equation}
S^{eff} = \frac{\int dE S(E)\frac{dN_{\nu}}{dE_{\nu}}}{\int dE
  \frac{dN_{\nu}}{dE_{\nu}}} 
\end{equation}
within energy range from the neutrino energy threshold $E_{th}$
to the DM mass $m_{DM}$ and with implied sum over neutrino and
antineutrino contributions. 
The neutrino effective area $S^i(E_{th},E_{\nu})$ of the NT200 for a
given configuration is a product of the efficiency
$\epsilon^{i}_{\nu}$ of muon reconstruction i.e. the ratio of
two-dimensional angular-energy distributions of reconstructed events
to simulated neutrinos, and neutrino impact area defined by MC
generated volumes $V^{i}_{MC}$ and the length of charged current
(CC) neutrino interactions in water depending on neutrino energy
$E_{\nu}$: 
\begin{equation}
\label{eqn:Seffnu}
S^i(E_{th}, E_{\nu},d\Omega) =  V^{i}_{MC} \times N_A \times \rho
\times \sigma^{CC}(E_{\nu}) \times  \epsilon^{i}_{\nu}(E_{th},
E_{\nu},d\Omega),  
\end{equation}
where  $i=1, ...,12$ refers to twelve configurations of the NT200.
During the livetime (between April of 1998 and February of 2003), the
NT200 took data in various configurations, which had been changed
  due to failure of some groups of OMs~\cite{Baikal-APJ2006}. Neglecting 
  few-OM differences, the data can be grouped according to twelve 
  configurations of the telescope. The values of corresponding
  effective areas $S^i(E_{th},E_{\nu})$ are varied within 25\% about
  the total effective area derived by averaging of
  $S^i(E_{th},E_{\nu})$ with the fractions of data taken period in
  particular $i$-th configuration of the telescope. The weighted
  average value of  the effective area is used in further analysis. 
The efficiency $\epsilon^{i}_{\nu}(E_{th},E_{\nu},d\Omega)$
is convoluted with the visible zenith track of the GC and thus
$S(E_{th},E_{\nu})$ is entering into Eq.(\ref{eqn:Seffnu}). The mean
value of the generated volumes is $V_{MC}=4.406\times{10}^{14}$cm$^3$.  
The value $N_A$ is the Avogadro number, $\rho$ is the medium
density (rock or water), ${\sigma^{CC}}$  is the
neutrino-nucleon cross section in the CC interactions. We omit the
exponential attenuation of the neutrino flux in the Earth in
Eq.(\ref{eqn:Seffnu}) since the shadowing effect is very weak for
energies less than 10 TeV.  
\begin{figure}[!htb]
\begin{center}
\includegraphics[width=0.35\textwidth,angle=-90]{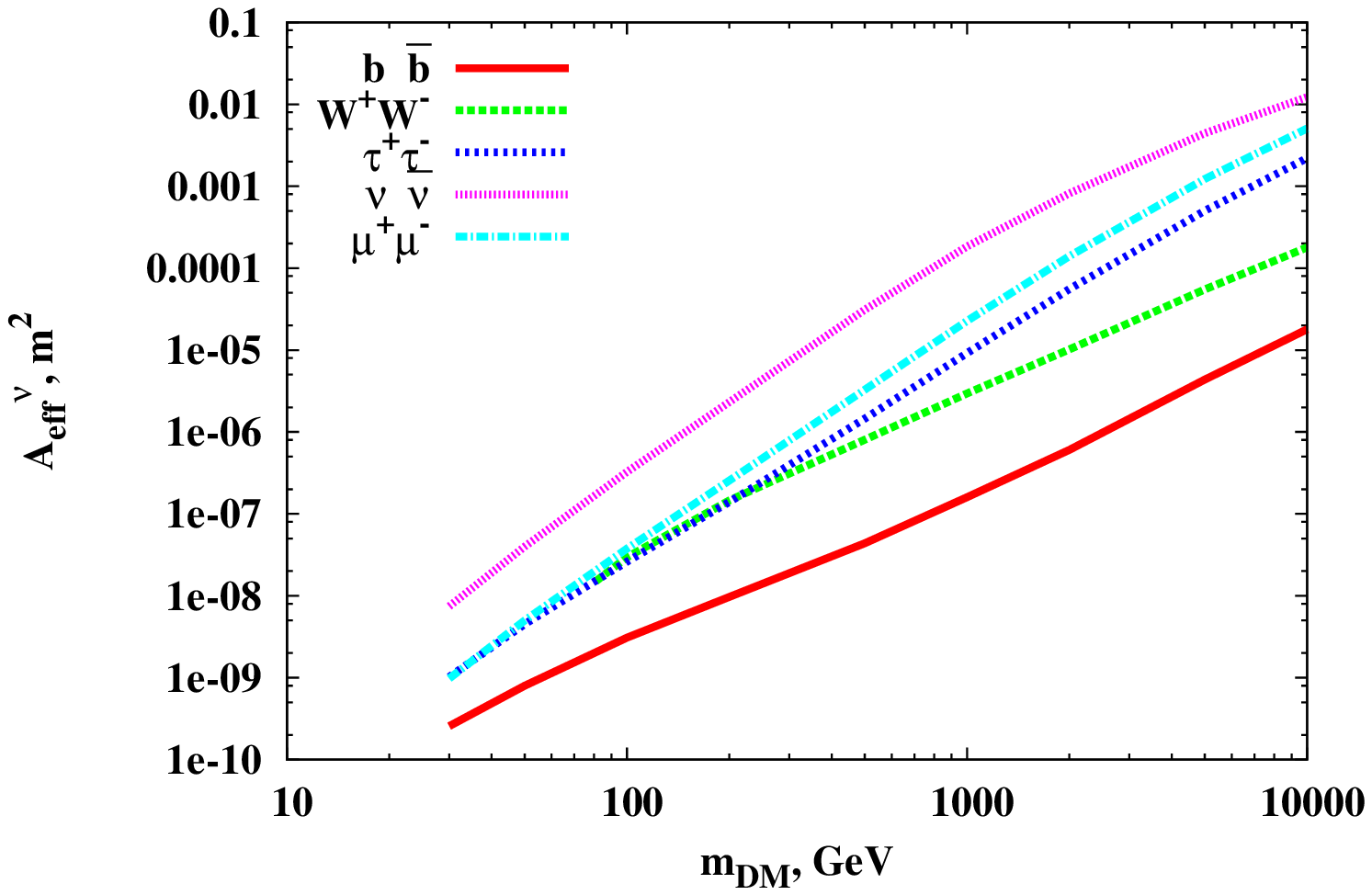}
\end{center}
\caption{\label{fig:6} 
The effective area as a function of the mass $m_{DM}$ of DM
particle for different annihilation channels.}
\end{figure}
The NT200 effective areas versus masses of dark matter particle for
particular annihilation channels are shown in  Fig.~\ref{fig:6}.
The softest neutrino spectrum in $b\bar{b}$ channel has the lowest
effective area, while the largest area is for hard pure neutrino
channel. The quantity $J_{a,\Delta\Omega}$ {in Eq.~(\ref{eqn:Seffnu})}
is obtained by   integrating the 
astrophysical factor $J_a(\psi)$ over the search region with
visibility $\epsilon(\psi,\phi)$ as follows
\begin{equation}
J_{a, \Delta\Omega} = \int d(cos{\psi})d\phi J_{a}(\psi)\epsilon(\psi,\phi).
\end{equation}
The upper limits for the dark matter annihilation cross section
are set by inverting the formula~(\ref{eq:6}) with respect to
$\langle\sigma_a v\rangle$ for a given annihilation channel. 
Obtained results with included systematic uncertainties are shown
in Fig.~\ref{fig:11} by dashed lines in comparison with those obtained
with method {\it B} (see next subsection).

\subsection{Method B: Maximum likelihood ratio}
In the analysis {\it B}, we construct likelihood function and use more 
detailed information about angular distributions of signal and
background. In this case, we choose sufficiently large signal region
$\psi< 40^\circ$  for counting of the events and disregard
remaining part as it is expected to be dominated by the
background. Let $f_S(\psi)$ and 
$f_B(\psi)$ be expected signal and background angular distribution
functions normalized in the search region. In the case of
nonzero number of signal events, one expects the data will be
distributed according to
\begin{equation}
f(\psi,N_S,N_B) = \frac{1}{N_S+N_B}\left(N_Sf_S(\psi) +
N_Bf_B(\psi)\right). 
\end{equation}
The likelihood function is then constructed as a product of
probability distributions for obtained events 
\begin{equation}
{\cal L}(N_S) = \frac{(N_B+N_S)^n}{n!}{\rm e}^{-(N_B+N_S)}
\prod_{i=1}^{n} f(\psi_i,N_B,N_S)
\end{equation}
with a multiplier with the form of the Poisson distribution
accounting for 
fluctuations in the total number of events. Here $n$ is the
number of 
observed events in the search region. Systematic uncertainties 
  for both signal and background are incorporated in the likelihood 
function as the nuisance parameters
$\theta\equiv\{\epsilon_S,\epsilon_B\}$ which are modeled by the
Gaussian distributions. They modify the probability likelihood function
as follows 
\begin{equation}
{\cal L}(N_S,\epsilon_S,\epsilon_B) =
{\cal N}\frac{(\epsilon_BN_B+\epsilon_SN_S)^n}{n!}{\rm
   e}^{-(\epsilon_BN_B+\epsilon_SN_S)-\frac{(\epsilon_S-1)^2}{2\sigma_S^2} 
-   \frac{(\epsilon_B-1)^2}{2\sigma_B^2}}
\prod_{i=1}^{n}
f(\psi_i,\epsilon_BN_B,\epsilon_SN_S) ,
\end{equation}
where $\sigma_S,\sigma_B$ are the systematic uncertainties which
will be 
discussed in the next Section and ${\cal N}$ is a normalization
factor.
Then,  the following profile likelihood 
\begin{equation}
\lambda(N_S) = -2\;{\rm ln}\frac{{\cal L}(N_S,
  \hat{\hat{\theta}}(N_S))}{{\cal L}(\hat{N_S}, 
  \hat{\theta})}
\end{equation}
is used to obtain upper limits on number of signal events. Here 
$\hat{N_S}$ and $\hat{\theta}$ are the values which give
absolute maximum to the likelihood probability function, while
$\hat{\hat{\theta}}(N_S)$ denotes the value of $\theta$ in the
  maximum of the likelihood at fixed value of $N_S$. The number
of observed events inside the search region $\psi< 40$ equals 113,
that is sufficiently large for the quantity $\lambda(N_S)$ to be
distributed according to the $\chi^2$ distribution with one 
degree of freedom according to Wilks
theorem~\cite{Wilks:1938dza,Agashe:2014kda}. We have checked this
numerically by running 10000 pseudo-experiments. Then we obtain
the upper limits on the number of signal events in the search
region at 90\% CL  by solving the equation
$\lambda(N_S^{90})=2.71$~\cite{Agashe:2014kda}.   

To set the upper limits on the dark matter annihilation cross
section for a particular channel and  DM mass, we use the same
inversion of the formula~(\ref{eq:6}) as in the description of the
analysis~{\it A}.   
Numerical values of the upper limits on the annihilation cross section
obtained at 90\% CL with the method~{\it B} are presented in
Table~\ref{tab:2} and in Fig.~\ref{fig:11} (solid lines).
\begin{table}[!htb]
\begin{center}
\begin{tabular}{|c|c|c|c|c|c|}
\hline
$m_{DM}$, GeV & \multicolumn{5}{c|}{$\langle \sigma_A v\rangle$,
  10$^{-21}$cm$^3/$s, NFW, method {\it B}} \\ 
\hline
& $b\bar{b}$ & $W^+W^-$ & $\tau^+\tau^-$ & $\mu^+\mu^-$ &
$\nu\bar{\nu}$ \\
\hline
30 & 8770 & --   & 76.9 & 69.3 & 5.86\\ 
\hline
50 & 1660 & --   & 25.4 & 21.9 & 2.88\\
\hline
100 & 392 & 20.9 & 9.61 & 8.21 & 1.35\\
\hline
200 & 144 & 8.19 & 4.71 & 4.04 & 0.742\\
\hline
500 & 60.9 & 3.35 & 2.01 & 1.72 & 0.324 \\
\hline
1000 & 31.6 & 1.91 & 1.09 & 0.957 & 0.201\\
\hline
2000 & 17.5 & 1.26 & 0.659 & 0.57 & 0.135\\
\hline
5000 & 8.55 & 0.926 & 0.414 & 0.364 & 0.111\\
\hline
10000 & 6.28 & 0.852 & 0.349 & 0.313 & 0.120\\
\hline
\end{tabular}
\caption{\label{tab:2} 90\% CL upper limits on annihilation cross
  section for the NFW dark matter density profiles; analyses {\it B}}
\end{center} 
\end{table}

\section{Results and discussion}
\begin{figure}[!htb]
\begin{center}
\begin{tabular}{cc}
\includegraphics[width=0.33\textwidth,angle=-90]{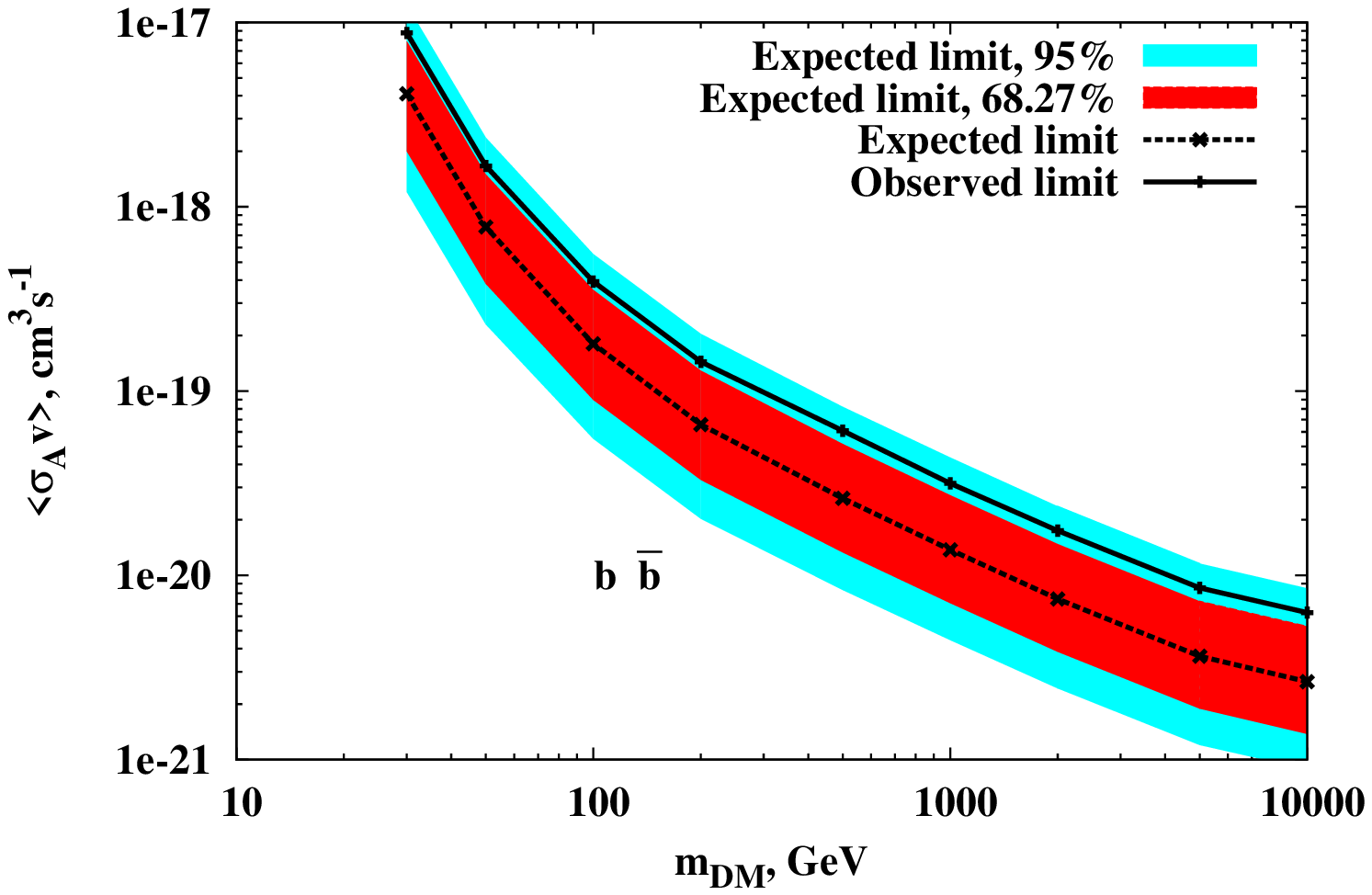}
&
\includegraphics[width=0.33\textwidth,angle=-90]{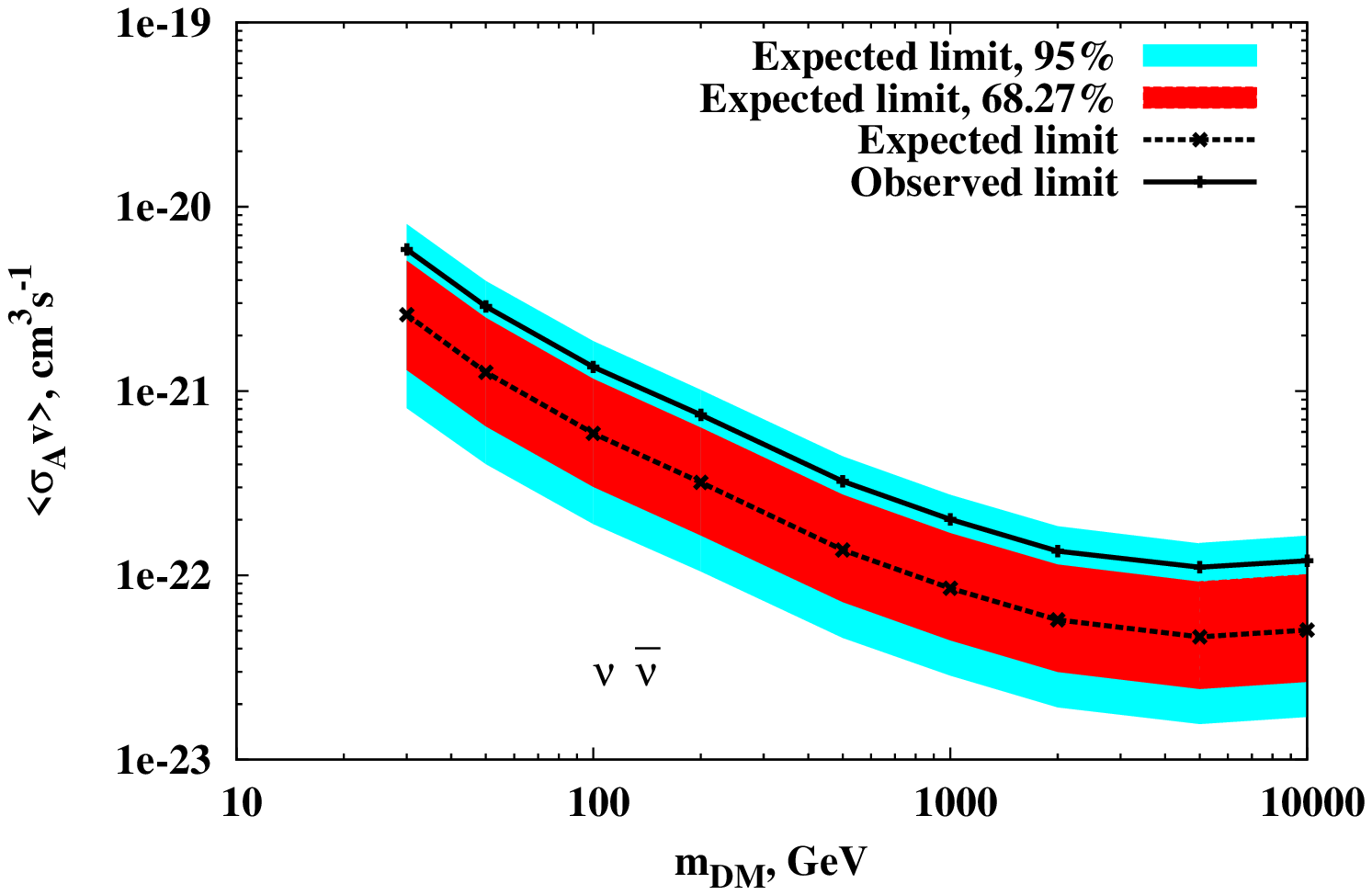}
\end{tabular}
\end{center}
\caption{\label{fig:7} 
The upper limits on the annihilation cross section versus DM mass
for channels $b\bar{b}$ (left) and $\nu\bar{\nu}$ (right) at 90\% CL
(black, solid) and the expected sensitivities  (black, dashed) within
its 1$\sigma$ (red band) and 2$\sigma$ (blue band) levels of
statistical uncertainty.} 
\end{figure}

\begin{figure}[!htb]
\begin{center}
\includegraphics[width=0.35\textwidth,angle=-90]{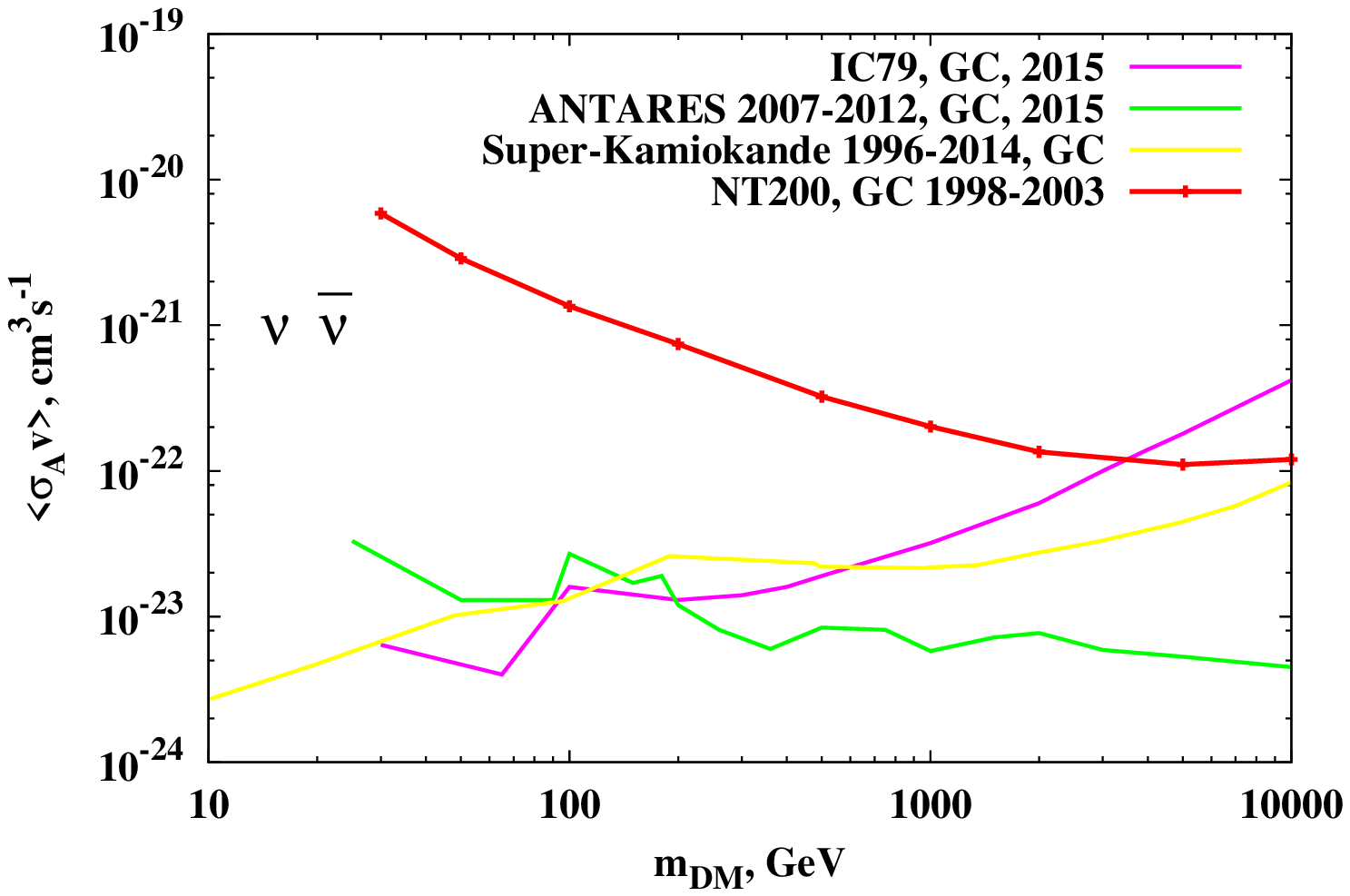}
\end{center}
\caption{\label{fig:8} 
90\% CL upper limits on the dark matter annihilation cross
section for $\nu\bar{\nu}$ channel obtained with the NT200
dataset (red line) in this study in comparison with results from
ANTARES~\cite{Adrian-Martinez:2015wey}, IceCube~\cite{Aartsen:2015xej}
and Super-Kamiokande~\cite{Frankiewicz:2015zma}. 
}     
\end{figure}

\begin{figure}[!htb]
\begin{center}
\includegraphics[width=0.5\textwidth,angle=-90]{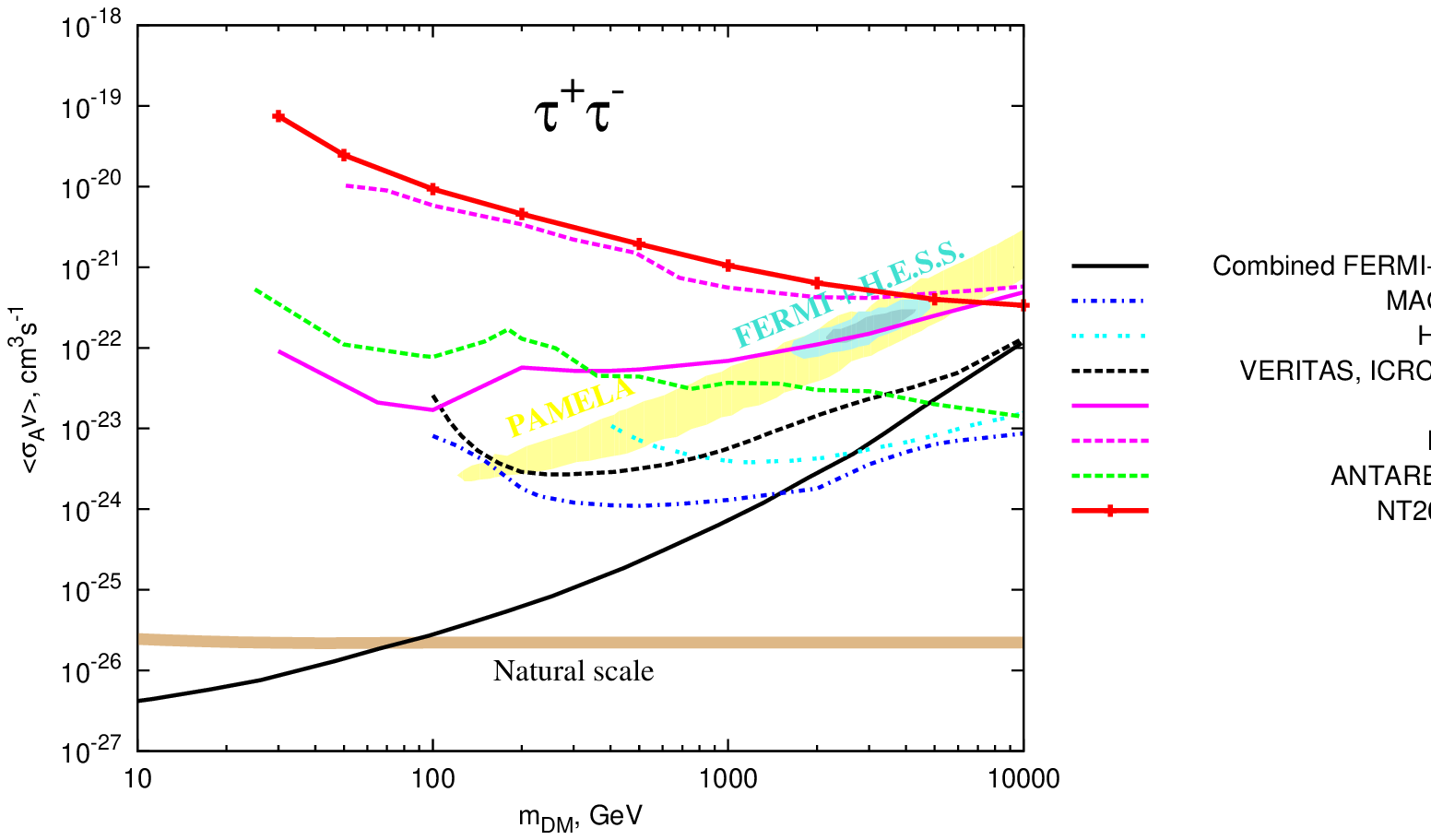}
\end{center}
\caption{\label{fig:9} 
90\% CL upper limits on the dark matter annihilation cross
section for $\tau^+\tau^-$ channels obtained from the NT200 data
in this study in comparison with results.}     
\end{figure}

We summarize our results for annihilation cross section for different
channels in plots shown in Figs.~\ref{fig:11}--\ref{fig:9}
comparing limits, obtained using methods A and B, with other
experiments. Firstly, we have obtained the consistent results of both 
analyses. In 
Fig.~\ref{fig:11} the upper limits at 90\% CL on the
annihilation cross sections of a DM particle for five particular 
channels obtained with the cone half-angle analysis (dashed
lines) and with the method of maximum likelihood ratio (solid lines)
are shown. Somewhat stronger upper limits are obtained with the
likelihood analysis for most of the chosen dark matter
masses and annihilation channels.   

We run pseudo-experiments with expected background only distribution
to estimate the NT200 sensitivity to the DM signal in the GC
direction for the opposite cases of soft $b\bar{b}$ and hard
$\nu\bar{\nu}$ neutrino spectra. The expected sensitivities at
90\% CL are presented in Fig.~\ref{fig:7} together with the upper
limits obtained by method {\it B}. 
Colored bands in these Figures represent 68\% (red) and 95\% (blue) 
quantiles. 
The observed upper limits are weaker as compared to the mean 
values of the sensitivity, because of a small excess of events 
in the direction towards the GC which will be discussed later in the
Section. 

Neutrino telescopes are most sensitive to the pure neutrino
  annihilation channels. In Fig.~\ref{fig:8}
we show the NT200 results (red line) for $\nu\bar{\nu}$ channel
along with the limits on the dark matter annihilations in
  the GC from other neutrino experiments, the
ANTARES~\cite{Adrian-Martinez:2015wey}, IceCube\cite{Aartsen:2015xej},
Super-Kamiokande~\cite{Frankiewicz:2015zma}. 
   We see that in accordance with the reconstruction efficiency for 
  sub-TeV neutrinos the NT200 upper limits obtained for $\nu\bar{\nu}$
  channel are weaker for smaller masses of the dark matter particles than 
  those in the range of masses above TeV scale. In the latter case,
  they are comparable with the limits from the IceCube (analysis based
  on the contained events) and the Super-Kamiokande. Obviously, the low
  energy part of neutrino events, which are upgoing through the NT200,
  is difficult to distinguish from the background of atmospheric muons
  and it is strongly rejected by the quality cuts. 

Other annihilation channels can be probed also by
gamma-telescopes\footnote{Here we note, that dark matter
  annihilations into monochromatic neutrinos are always produced with
  some amount of photons generated by electroweak
  bremsstrahlung. Thus, $\nu\bar{\nu}$ channel can be indirectly
  probed with gamma-ray telescopes, see recent discussion
  in~\cite{Queiroz:2016zwd}.}.  
In Fig.~\ref{fig:9} we compare upper limits at 90\% CL on the
$\tau^+\tau^-$ annihilation cross section obtained with the
NT200 dataset and other experiments including
results by the FERMI~\cite{Drlica-Wagner:2015xua} (dwarf
galaxies, DES), VERITAS~\cite{Zitzer:2015eqa} (four dwarf galaxies),
MAGIC~\cite{Aleksic:2013xea}  (Segue~1),
HESS~\cite{Abramowski:2014tra} (dwarf  galaxies) ,
IceCube~\cite{Aartsen:2015xej,Aartsen:2015bwa} (GC and
  preliminary results for dwarf galaxies) ,
ANTARES~\cite{Adrian-Martinez:2015wey} (GC). Colorful regions
show the results of DM interpretation~\cite{Meade:2009iu} of positron 
excess observed by PAMELA~\cite{Adriani:2008zr} combined with
the FERMI~\cite{Abdo:2009zk} and HESS~\cite{Aharonian:2008aa}
data.  Light brown line shows the thermal relic annihilation cross
section from  Ref.~\cite{Steigman:2012nb}. 

\begin{figure}[!htb]
\begin{center}
\includegraphics[width=0.35\textwidth,angle=-90]{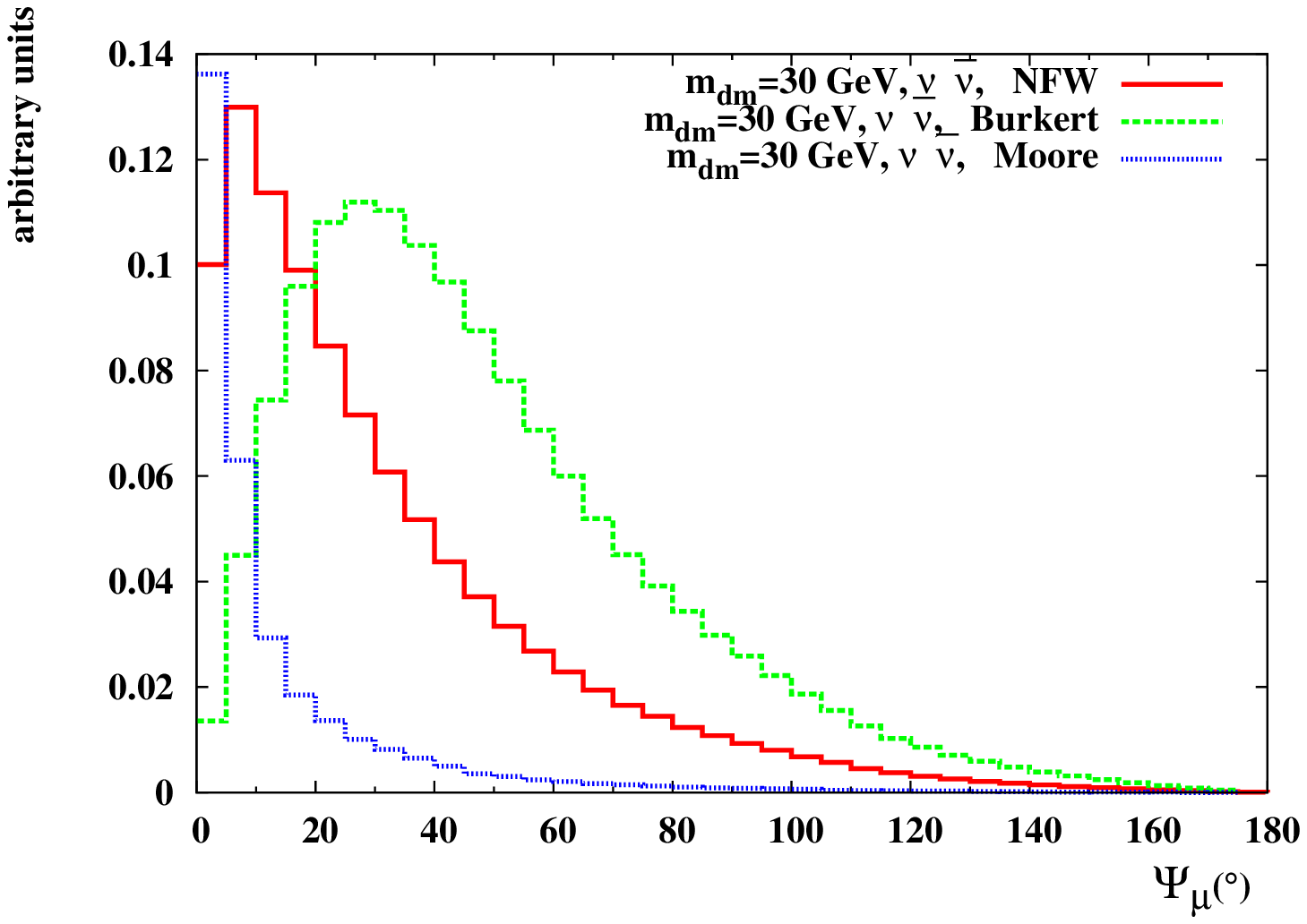}
\end{center}
\caption{\label{fig:12} 
Reconstructed muon angular distribution from the dark matter annihilations 
over $\nu\bar{\nu}$ channels for different dark matter density 
profiles. Relative normalizations are scaled for convenience of
presentation. }     
\end{figure}

\begin{figure}[!htb]
\begin{center}
\includegraphics[width=0.35\textwidth,angle=-90]{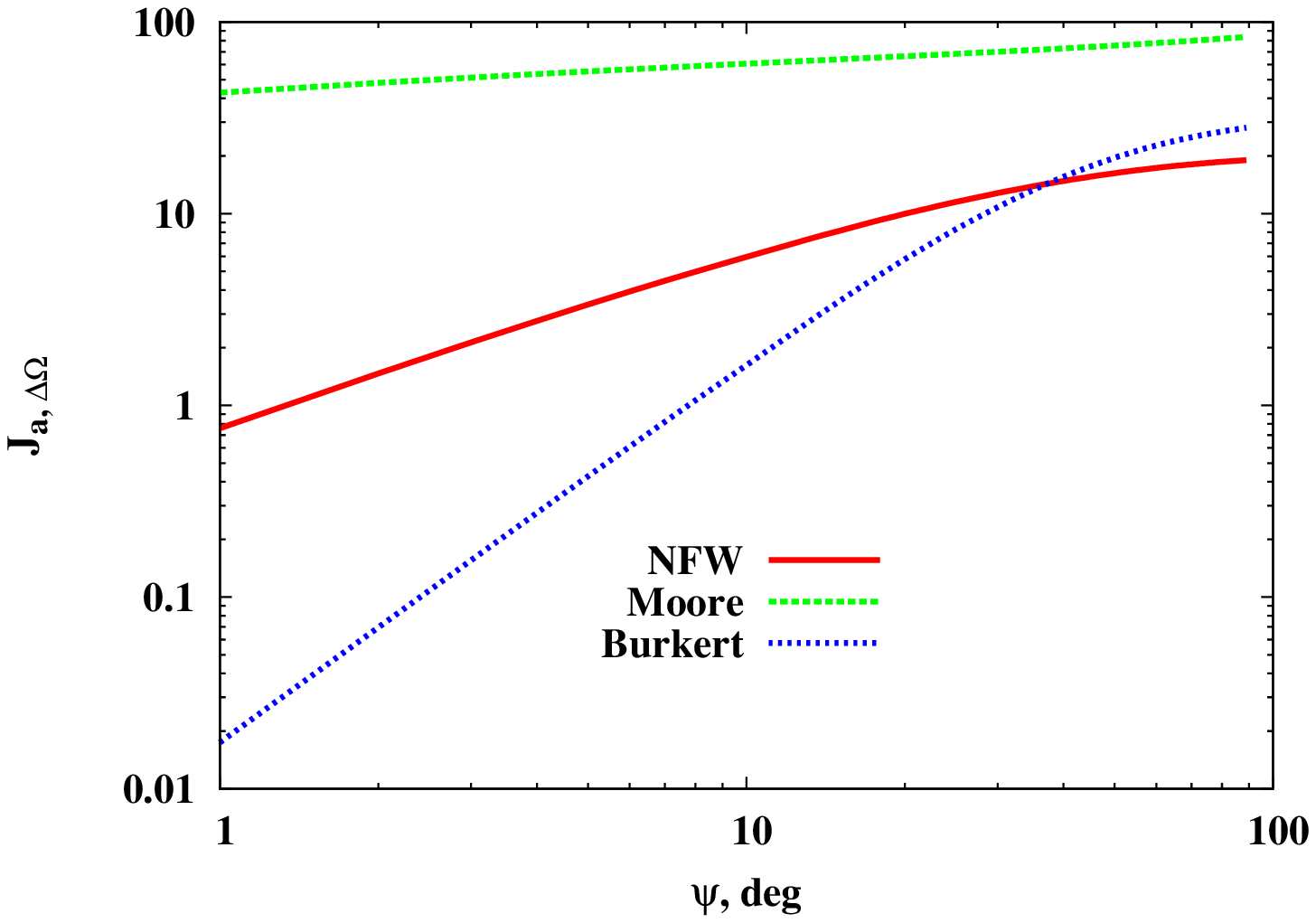}
\end{center}
\caption{\label{fig:13} 
$J_{a,\Delta\Omega}$ as functions of the cone half-angle $\psi$ for
  different matter density profiles of the Milky Way.}     
\end{figure}

\begin{figure}[!htb]
\begin{center}
\includegraphics[width=0.35\textwidth,angle=-90]{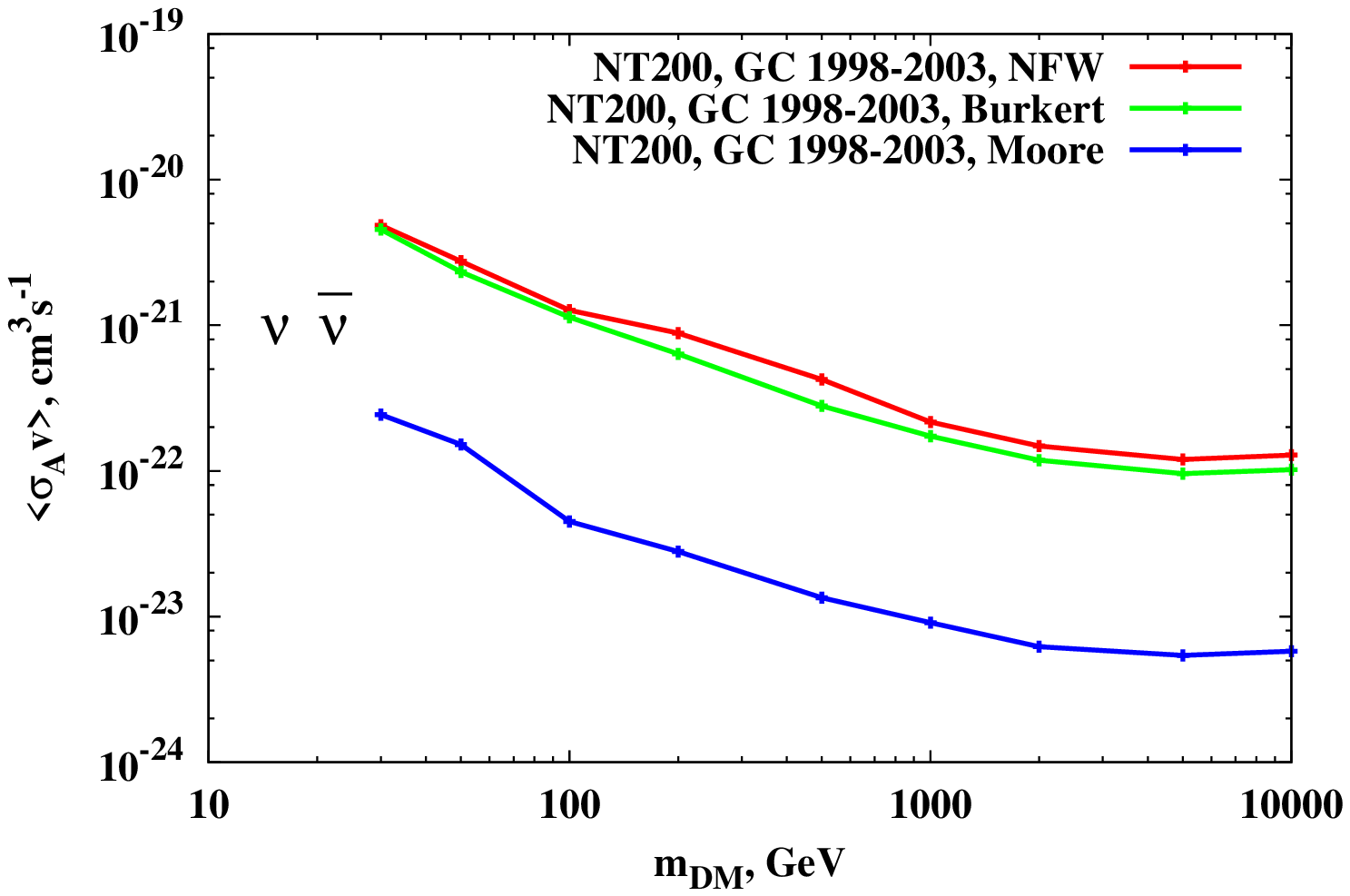}
\end{center}
\caption{\label{fig:10} 
90\% CL upper limits on the dark matter annihilation cross
section for $\nu\bar{\nu}$ channel obtained from the NT200 data
in this study for different profiles of dark matter density in the
Milky Way.} 
\end{figure}

Let us further discuss the systematic uncertainties. They include
both experimental and theoretical parts. The uncertainties in the
optical properties of water and in the sensitivity of the optical
modules result in 30\% experimental
uncertainty~\cite{NT200Mono-2008, NT200Astro-2011}. 
The variations of the effective area related to various telescope 
configurations is estimated to contribute the systematic error
less than 1\%.
Theoretical uncertainties include present errors in
the oscillation parameters and uncertainty in the
neutrino-nucleon cross sections. They have been 
estimated using the procedure described in~\cite{NT200Sun:2014swy} and
reach 10-12\% depending on annihilation channel. We include the
above uncertainties when presenting the upper limits. However,
the main uncertainty comes from astrophysics. To illustrate the
  influence of the astrophysical uncertainty on the upper limits on
  the dark matter annihilation cross section, we
carry out new analysis using the method~{\it A} with the Burkert and
Moore dark matter density profiles for the monochromatic neutrino
  annihilation channel: we make a new MC simulation of the signal
events and obtain new values of optimized cones. In 
Fig.~\ref{fig:12} 
we show for illustration expected reconstructed muon angular
distribution from this signal for $\nu\bar{\nu}$ annihilation channels
for different dark matter density profiles. Using optimization
procedure, we found that for the Burkert profile the cone
half-angle should  be about $56-57^\circ$ while for the Moore
profile we obtain values in 
the range $3-11^\circ$ depending on the mass of the dark matter
particle and annihilation channel. Corresponding integrated
$J_a$-factors as functions of cone size are shown in Fig.~\ref{fig:13} 
for chosen dark matter density profiles. The results for upper limits
on the annihilation cross section are presented in
Fig.~\ref{fig:10}.  
We see that for very cuspy Moore profile the bounds on the
annihilation cross section are improved by an order of 
  magnitude. This is related to considerably larger values of
$J_a$-factors even for 
very small opening angles as compared to the NFW profile. At the
same time the cored Burkert profile results in almost the same
upper limits. In this case, the large background from new opening
angle is compensated by larger astrophysical factor. We see that
$J_a$ for the case of the Burkert profile is larger than for
the  NFW case at $\psi\gsim37^\circ$. 

 Obviously, due to observed small excess of events towards the GC
   we obtain a weakening of the upper limits on the dark matter
   annihilation cross section as compared to sensitivity. Calculating
   the test statistics (TS) of this excess without any systematic errors
 under background only hypothesis we find values $5.8-6.6$ depending on
  annihilation channels. We perform a study of the influence of
  different systematic errors on the statistical significance of this
  excess. To consider astrophysical systematic uncertainty we use 
  the NFW profile but allow corresponding parameters $r_*$ and
  $\rho_*$ vary within the following $2\sigma$ bands
\begin{equation}
r_* = 16.1^{+17.0}_{-7.8}~{\rm kpc},\;\;\;
\rho_* = 0.533^{+1.104}_{-0.354}~{\rm GeV}/{\rm cm}^3
\end{equation}
taken from Ref.~\cite{Nesti:2013uwa}.  Their variations change
   the corresponding $J_a$-factor and thereby modify the
   signal both in the expected angular distribution and total flux. 
 We include $r_*$ and $\rho_*$ as
additional nuisance parameters in the probability likelihood function
similarly to $\epsilon_S$ in eq.(11) except that we use not Gaussian
but log-normal 
distribution to model this uncertainty. We have found no considerable
influence of the systematic uncertainty in signal (both from
astrophysics and particle physics sides) on the statistical
significance of the excess. However, inclusion of systematic
uncertainty in the background determination considerably brings down 
the values of TS to $1.4-1.6$. Thus, we attribute
this excess to a statistical fluctuation of the background only
expectation within less than 2$\sigma$ level.  

Finally, let us mention the uncertainty in $J_a$-factor related to a
possible asphericity of dark matter halo~\cite{Bernal:2014mmt} which
can reach values up to 10\%. We conservatively did not take
  this part of astrophysical uncertainty into account in our analysis.

\section{Conclusions}
\label{sec:conclusions}
To summarize, we studied the NT200 response to neutrinos
from dark matter annihilations in the Galactic Center. We 
  performed two independent analyses looking for an excess of
neutrino events towards  
the GC. Both analyses show similar results. The upper limits on 
  the dark
matter annihilation cross section for 2.76~live years of observation
reach values of about $10^{-22}$~cm$^3$s$^{-1}$ at 90\% CL
in channel of $\nu\bar{\nu}$ pairs when mass of the dark matter
particles is heavier than 5 TeV. 
We studied influence of the uncertainties related to
the dark matter density profile on the upper limits and found that 
they can result in a change of order of magnitude.

Finally, we remark that this study of neutrino signal from the GC
  was performed  for the first time by the Baikal collaboration and we
  consider it as an important step towards the future analysis of the
  data of the ongoing GVD-project~\cite{GVD2015a}.

\section*{Acknowledgments}

{\footnotesize
The work of S.V.~Demidov and O.V.~Suvorova was supported by the RSCF
grant 14-12-01430.  
}   




\bibliographystyle{model6-num-names}
\bibliography{<your-bib-database>}



\end{document}